# ENTERPRISE MODEL LIBRARY FOR BUSINESS-IT-ALIGNMENT

Peter Hillmann, Diana Schnell, Harald Hagel and Andreas Karcher

Department of Computer Science,
Universität der Bundeswehr, Munich, Germany

## ABSTRACT

*The knowledge of the world is passed on through libraries. Accordingly, domain expertise and experiences should also be transferred within an enterprise by a knowledge base. Therefore, models are an established medium to describe good practices for complex systems, processes, and interconnections. However, there is no structured and detailed approach for a design of an enterprise model library. The objective of this work is the reference architecture of a repository for models with function of reuse. It includes the design of the data structure for filing, the processes for administration and possibilities for usage. Our approach enables consistent mapping of requirements into models via meta-data attributes. Furthermore, the adaptation of reference architectures in specific use cases as well as a reconciliation of interrelationships is enabled. A case study with industry demonstrates the practical benefits of reusing work already done. It provides an organization with systematic access to specifications, standards and guidelines. Thus, further development is accelerated and supported in a structured manner, while complexity remains controllable. The presented approach enriches various enterprise architecture frameworks. It provides benefits for development based on models.*

## 1. INTRODUCTION

Over the years, companies and their employees aggregate a lot of expertise and experience. This ranges from structured processes to technical skills. Explicit expert knowledge can be well mapped by documentation. In contrast, implicit experience is difficult to explain or write down [1, 2, 3]. Furthermore, the totality of all information with its variety and diversity can lead to information overload and in-transparency. This problem is addressed by the approaches of model formation and *Enterprise Architecture* (EA), see Figure 1. Thereby, the central element is the Enterprise Architecture Library (EAL), in which the information is structured and manageable. Despite capability-based frameworks like *The Open Group Architecture Framework* (TOGAF), *IT Infrastructure Library* (ITIL), and *NATO Architecture Framework* (NAF), there is no practical template or content structure of an EAL. According to the definition, an EAL organizes models with reference character and user application according to uniform aspects [4]. It allows a systematic access on the knowledge of the business specific perspectives with e. g. value chain, processes, methods, applications and resources. In addition, a holistic EA landscape includes a comprehensive method and strategic orientation to the application of the models.

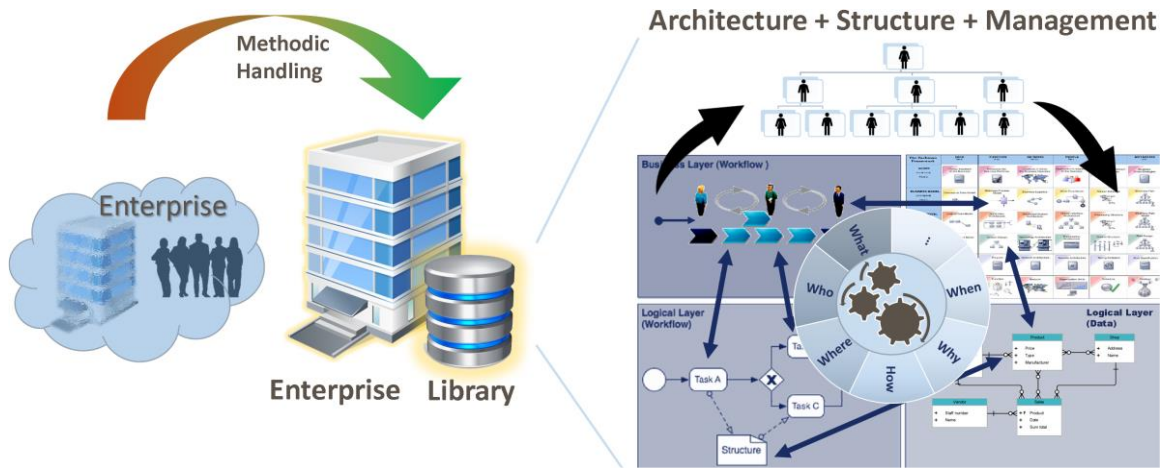

Figure 1. Reference design for an EAL of enterprise models and their maintenance

Thereby, the facts of an enterprise are illustrated via models. It is an adequate medium to describe *good practices* for complex systems, processes, and interconnections. Models provide a clear overview of interrelationships and allows them to be managed. Furthermore, the models have the goal of representing responsibilities, established methods and valid solutions. Thus, the broad knowledge of various company divisions become structured and made available for communication as well as reuse. Considering, especially reference models provide support with a recommendation character [5]. In the design process, best practices such as ITIL, COBIT, or IT4IT frameworks also serve as a guide for the modeler. These frameworks can be part of a company's own repertoire of reference materials. Therefore, a catalog organizes this information according to be defined criteria. According to Kiehl [6], the use of reference models shortens the development times of concrete models, while their model quality increases and the costs of modeling are reduced. Furthermore, it is hardly possible to feed back knowledge gained during the practical usage of a model [7].

To drive an enterprise forward, well-functioning Business-IT-Alignment is indispensable today. A central EAL of business models represents the possibility to counteract the information intransparency and discrepancy. Currently, there is no structured and detailed approach for the design of an EA model repository. In this paper, an EAL for models is developed to capture the wealth of information about an enterprise. It includes the reference design of the data structure for managing, processes for administration and possibilities for usage. In addition, our approach supports the maintenance of EA models. Thereby, relationships between entries, models, and architecture building blocks are shown. The EAL allows a targeted application and re-usability of already coordinated knowledge. To include an improvement possibility, the EAL supports the process with additional meta information and feedback options to an entry. So, it forms a knowledge base for the enterprise and their value streams. The structured preparation of established EA models is essential for it and influences the retrieve-ability of the generally valid solutions. It allows systematic access to existing knowledge, especially for new employees, special cases, and audits. The operation of an EAL supports the strategic management of the business. Generally valid solutions in the corporate environment are accordingly reusable to avoid work already done.

For clarification, we see a repository in this context as an unstructured collection of EA models. In contrast, an EAL provides structured access to content via a catalog as well as further functionality for application and usage.

This paper is structured as follows: In Section 2 we describe a typical scenario and the requirements for a reference EAL. Section 3 provides an overview of the current state of the art in that application area. The main part in Section 4 describes our concept of a reference EAL for models. Subsequently, we elaborate first experiences on prototypes and evaluate fundamental properties of the presented system in Section 5. The last section summarizes our work and provides an outlook.

## 2. SCENARIO AND REQUIREMENTS

The need for an easy-to-use EAL is illustrated by the following typical scenario. *YOUrRobots* is a mid-sized company and offers autonomous robots for the maintenance of green areas. TOGAF is already established to cyclic improve the interaction between IT infrastructure and the implementation of business goals. According to model-driven development, the technical models and documents can be reused as reference other business areas with the knowledge already gained. The methods and processes are harmonized for complexity reduction and integrated Business-IT-Alignment.

To increase customer satisfaction, new robots are to be equipped with additional services by plugin extensions. The smartification includes remote administration for planning and control as well as online software updates. In addition, it offers a regular evaluation of usage data for maintenance depending on actual wear. Therefore, the company will operate several cloud services for remote administration by customers. In order to reduce the maintenance effort and the operational costs for setting up services, their structure should be identical. Appropriate referral models offer flexibility and support integration into existing structures. Also, the management requires support in the IT alignment of the business units. Therefore, the introduction of an EAL is pushed with the re-usability of the adapted solutions and strategic management.

The challenge is to provide the knowledge gained in a structured way, especially for new employed systems engineers. In addition, the aligned relationships between reference approaches and adapted solutions need to be supported. For cooperation of different departments, the information has to be interconnected with adequate interfaces. The diverse structures and processes require various types of data, from simple descriptions to visual models and complex specifications.

Generally, most important requirements for an EAL and its contents are the following:

- **R1**: The EAL has to provide sufficient documentation of model purpose, model context, problem structure and design aspects, so that communication is improved and interpretation is limited.
- **R2**: The EAL shall support storage and re-usability of architecture components of different levels to simplify the development of new models.
- **R3**: The approach shall provide discoverability via typology and feature search, factual logical access hierarchies, and classification by feature description.
- **R4**: The EAL entries shall have a recommendation character and therefore be configurable, adaptable, instantiable and extensible.
- **R5**: The approach shall support linkage between versions and variants of EA components and models built on top from them.

Based on the mentioned requirements, we consider the following research questions to be important:

- **Q1**: How can an EAL be realized to support a sustainable approach through reuse of architecture components?
- **Q2**: What properties must an EA component have in order to have reference character and actually be reusable?
- **Q3**: How can interrelationships and dependencies between components be made available to support and guide a modeler?
- **Q4**: What is the process for integrated application of models via library throughout a life-cycle?

## 3. RELATED WORK

The requirement for reference modeling has existed since 1980 [8]. Due to the diversity, the need for a repository has also arisen. The term reference comes from etymological and means recommendation. According to this, a reference model has recommendation character or it is referenced [9].

According to ISO/IEC 42020 definition, the architecture repository hosts the baselines of architecture elements produced or updated by architects. It includes different kinds of architectures models and architecture elements like patterns and building blocks. It is a place where work products and the associated information items can be stored for preservation and retrieval. In alignment with the TOGAF enterprise continuum, it describes elements with increasing detail and specialization.

Various references to an architecture repository can be found in the literature. In the context of the EA Framework TOGAF, the consortium *The Open Group* roughly describes an EA repository with its components [10]. It contains out of six main classes of architectural information: Architecture Meta-model, Architecture Capability, Architecture Landscape, Standards Information Base, Governance Log, and Reference Library. In addition, the concept includes connections to a *Solution and Requirements Repository* for a suitable interplay in the application domain. Based on the mere existence of such a repository, its added value and usefulness are made clear. The main comparative component in relation to our scenario is the *Reference Library*. However, the description is an empty shell with a superficial description of the possible content. A detailed description is missing regarding the structuring, process and procedure of application. According to given templates of *The Open Group*, only a folder structure on file basis is suggested [11]. ArchiMate also recommends a file-based storage in the scenario for banking industry [12]. This is also accompanied by a specific model registry for logistic [13].

The newer approach of the NATO Architecture Framework (NAF) describes a comparable reference library with the Architecture Landscape [14]. This contains a basic set of assets that can be reused by architects. Lang [15] distinguishes here three kinds of model libraries, in order to separate the conceptually different reference models. It is differentiated into the following libraries: domain-neutral, domain-specific, and enterprise-specific. A detailed description for a repository is not given.

For a corresponding classification and reuse of models, there are first approaches of a taxonomy according to application areas [16]. Another classification is based on structural and model aspects [17]. Both approaches are on a theoretical level. In addition, the rough methodical procedure for reference modeling are given with the top-down or bottom-up strategies [18, 19]. Besides this, an abstract infrastructure for a model management system is illustrated. Furthermore, it should be noted that there are different types of reference modeling [20, 21]: Analogy Construct, Specialization, Aggregation, Instantiation, and Configuration. This results in three possibilities of construction for the application of models in the specific case: variants, versions, and re-construction.

In the ITIL approach, several databases with similar goals to a repository are described [22]. Their Service Knowledge Management System is intended to centrally manage the essential assets of a company. Nevertheless, no detailed description is provided.

In addition, the Gamma et al. [23] design patterns provide detailed modeling and descriptions of recommended solutions to typical problems in information technology. However, these do not relate to enterprise design and no repository structure is given.

Approaches for EA design can be found in the patterns of Perroud and Inversini [24]. Here, some patterns are described on the basis of an individual meta-model. Possible categorizations are provided for software and construction pattern, which can be partially adapted.

The comprehensive approach to a repository is provided by the German University of Saarbrücken [25]. It classifies many different models. Nevertheless, their reference model catalog focus on a tabular without individual adaptability and operational solution. According to our requirements it lacks structural coherence, graphical representation and practical usability.

The IT4IT framework clarifies adaptable information objects via the reference architecture [26]. According to the EA Functional Component data object, only the three attributes *ID*, *Component*, and *Diagram* are suggested.

In summary, there is so far no practical usable approach of an EAL for the management and usage of models in long-term.

## 4. CONCEPT OF A LIBRARY FOR ENTERPRISE MODELS

Our holistic approach to an EAL addresses these topics in detail with a domain-independent description of such a system. Figure 2 provides an overview of the main components. The focus is on the information and application management.

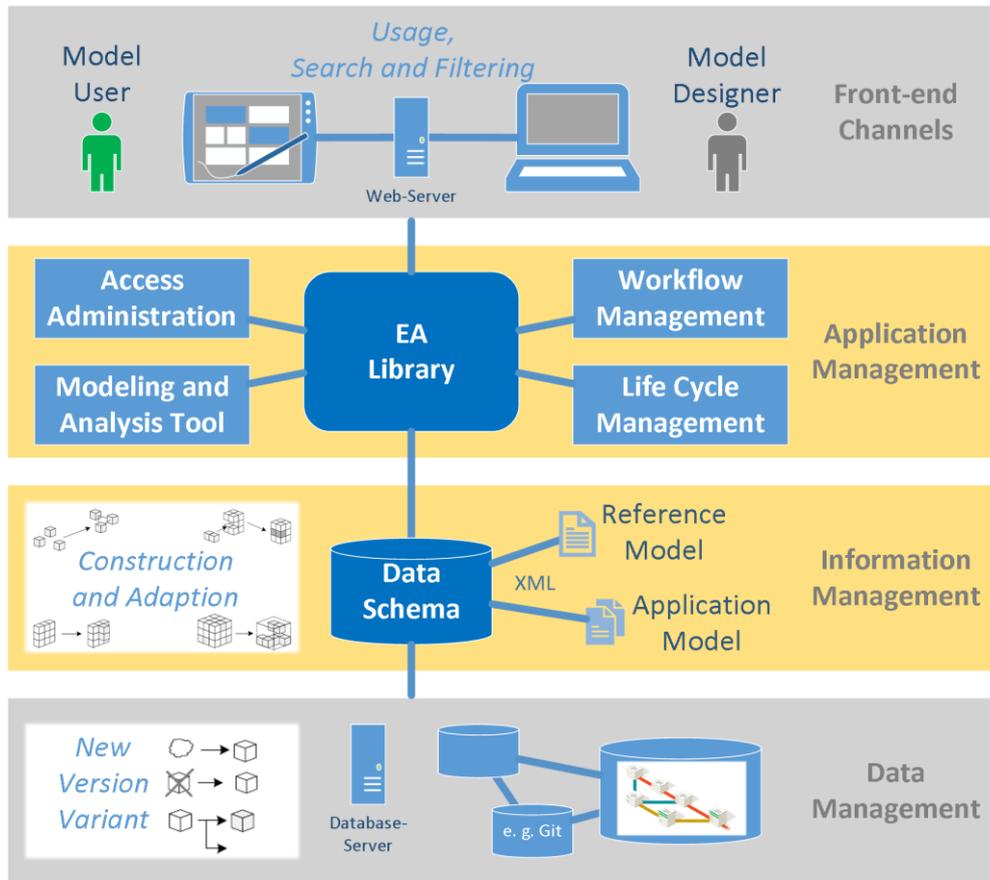

Figure 2. Overview of the Enterprise Architecture Library.

At the bottom line is the data management based on a vault for linking of versions and variants, see Section 4.2. It is setup on a database to manage the entries with meta-data, models, and optional content as well as access rules. A distributed file system is suitable here, which enables version and variant management with role-based access control, e.g. Git. This infrastructure supports collaborative work with non-linear workflows. Based on this, the structure for orderly information management is specified. Here, the storage of entries takes place in the form of a catalog, see Section 4.1. This includes the models with reference and application character. Furthermore, it offers architecture building blocks, which can be used for new constructions, further development, and flexible adaptation. For this, we use the standardized data format XML (based on *The Open Group ArchiMate Exchange File Format*).

The core of the EAL system lies in the application management. This consists mainly of four parts. The access administration manages the users with their roles and rights. It controls the access to the system and limits the modification of entries to responsive role. Here, the organizational structure can be used accordingly. The modeling and analysis tools support the designer during the construction and modeling. These depend on the individual needs and use case of the modelers and the company. Here, common meta-models with the favored tools are applied. In particular, building block-oriented modeling is an essential feature with regard to reusability. It also enables adaptation to a specific problem in a domain. For this purpose, the modeler must pay attention to clearly distinguishable building blocks. These must not be designed too specifically, especially in reference modeling, and must be provided with clearly defined interfaces.

The flow of information is controlled by means of workflow management, see Section 4.3. A process engine is used for this purpose, which provides defined processes for modeling, application and usage. The workflows are supported semi-automatically by stringently following the process model and generating attributes of the entries. This also allows company-specific workflows to be mapped, such as change proposals. In connection with workflow management, life-cycle management is carried out for the entries and models. Models of different levels and perspectives can be entered via the targeted attribution in the library. An entry in the library follows a life-cycle, whereas the status is traced via specific attributes. The objective is a single source of information, whereas data linkage is enabled. Thus, separate directories for different continuum's are not mandatory in relation to model's abstraction level or purpose.

On the top level, the system is accessible via a web front-end. This guarantees an interaction across all devices in any situation. In summary, this creates a system for documenting and processing the knowledge that exists in the company in a structured manner. This counteracts the current obstacle of limited communication and makes the scope for interpretation manageable.

## 4.1. Information and Data Management

The following attributes are specified to meet the requirements for a structured storage of models and other information. The integration of a model in the library contains generated, mandatory, and optional parts to be filled.

The following attributes are **automatically generated**:

- **ID**: Unique identifier.
- **Version**: Ascending Number with creation date.
- **Status**: Current phase in the life-cycle with date of change.
- **Complexity**: Generated rating for syntax comprehensiveness.
- **Connectivity**: Generated score for semantic understanding.

The ID is mandatory for unique identification and referencing, equivalent to IT4IT. The tracking of improvements is done via appropriate versioning with sequential succession. The complexity and connectivity express how difficult the model is to capture and applicable. So far, there is no measurable metric with regard to "good" models. Experience has shown that a 3-level rating for both yardsticks is enough in practice, based on the number of elements and connections. By analogy for the comprehensibility of sentences, a corresponding measure is adapted here [27, 28]. For complexity, we rate models based on the sum of elements and connections. A model is easy with less than 20 components, moderate from 20-40 components, and complex with more than 40 components. For connectivity, we score models based on the average number of connections of an element. We see a model as simple with a score less than 2, average from 2-3, and difficult over 3. If a model is too complex or difficult, then it is suggested to subdivide the model into several sub-models or to structure it hierarchically [29].

The following **core attributes** are the key information for any entry to be included in the library:

- **Title**: Short and concise name of the model for the first relevance decision.
- **Category**: Kind of model with regard to structured reuse according to a taxonomy.
- **Layer**: Recommended level of application based on an enterprise hierarchy.
- **Abstract**: Description of the data set including pattern intent:
  context, problem statement, solution, and results.
- **Keywords**: Supports finding according to a feature list based on taxonomy.

- **Responsible Authors**: Central point of contact for questions and improvements.
- **Model**: Main content of an entry, which provides a solution approach for an use case.

The category represents an abstract high-level tag and is independent of the application area. It focuses on the enterprise continuum with domain-neutral, domain-specific, and company-specific aspects. The kind of the model is described with regard to possible reuse, e. g. building block, design pattern or application model. Especially, the design of reference models and generic building blocks require clear distinctiveness from each other and well-defined interfaces. The assignment to a layer addresses the subject of interest and describes the degree of abstraction as well as the level of detail. The specification of possible layers is oriented on the modeling framework, e.g. ArchiMate from the top with strategy to the bottom with physical systems. Typical areas are, e.g. Business Process, Application Stack, or Support Information-flow. The abstract describes the essential addressed contents of the model in the form of a succinct text. The keywords support the targeted retrieval via search and filtering. The specification of possible keywords is based on ontology, thesaurus, or glossary. According to the rights and roles model, there is a responsible author for the data set. This is the central point of contact for questions. The person is responsible for continuous improvement as well as maintenance and care. The essential content of an entry is given by the model, regardless of whether it is a reference, a building block, or an application. Based on a meta model, a solution approach is shown for an issue. This is not only used for documentation, it is mainly intended for reuse and further development.

Beside the key attributes, we suggest to added the following **optional information**:

- **Application** Context: Explanation of the circumstances when and if it should be used.
- **Stakeholder**: Matrix of important people to be addressed.
- **Capabilities**: Textual description of how to use the model and what skills must be met.
- **Limitations**: Listing of restrictions, requirements and possible challenges in application.
- **Dependencies**: References and connections to other facts, esp. for future developments.
- **Bricks**: Index of applied building blocks in this entry for traceability.
- **Variants**: Links to domain- or case-dependent modification or adaption; parallel existence.
- **Example**: Description of a concrete use case with application context for typical usage.
- **References**: List of sources for further information.

## 4.2. Structure and Library Administration

A vault is provided for structured storage of the hierarchical data. Therefore, the generic approach of a *Product Data Management Enabler* is adapted [30]. The Figure 3 shows the generic data schema of the back-end system. On the top level of the left part, the initial entry is created by means of the data element *Entry Master*. This describes the type of information that is stored in the library. It includes the automatically generated information and the obligatory meta-data. These are intended to describe management information and offer a flexible reuse. Both versions and variants are attached to the *Entry Master*. The actual data is stored under the *Entry Data* for a specific variant in a particular version. By definition, these describe the content and structure. It includes textual and graphical descriptions, building blocks and associated components. In addition, the data object can be supplemented with further information. Each entry can be assigned optional data and conditions on qualification, effectiveness and possible alternatives.

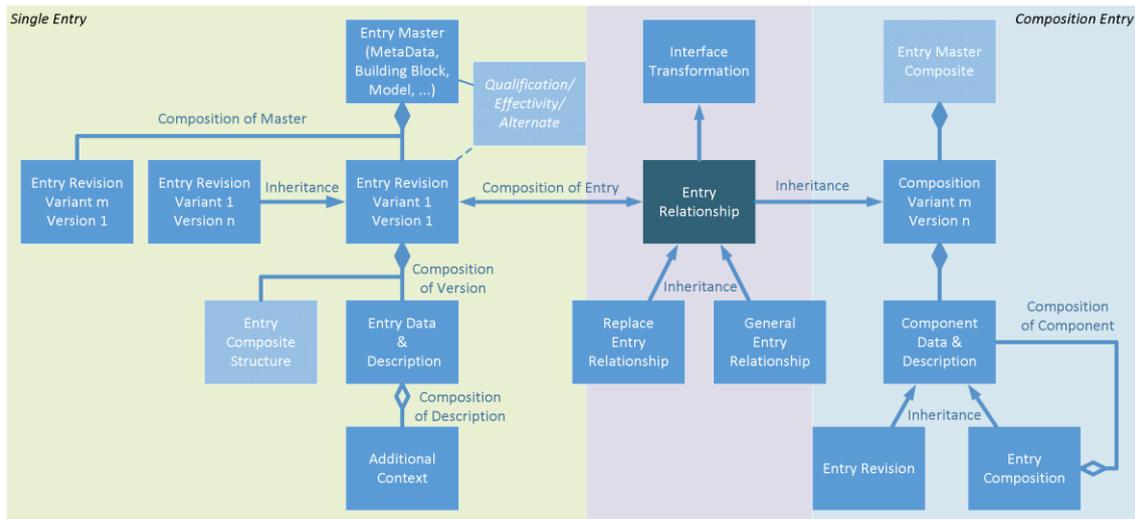

Figure 3. Hierarchical object-based data schema in the back-end of the vault.

For the reuse of models and building blocks a composition of entries is enabled, see right part of Figure 3. Analogous to the data structure of a general entry, the top element is called the master composite. This contains basic meta-data about the content. This is followed by the variant and version. According to the Design Pattern Composite a nesting of the data is made possible. A link to the original data record is established via a relationship so that no redundancies have to be maintained. A relation always refers to a specific variant and concrete version, so that possible incompatibilities in future versions must be checked in advance manually. The relation distinguishes between a general linkage and a replacement behavior. Depending on the compatibility, corresponding interfaces must be transformed.

## 4.3. Workflow of Usage and Customization

For the general use of the system, it provides an overview page with a grid structure according to aspects and subjects of the viewing angle, following [31]. So, it represents the different domains of the stored models and the content is grouped according to the taxonomic categorization [32]. In addition to a clear entry point, it offers easy accessibility. This architecture landscape already provides information about existing and open areas. In addition, there is the possibility of a search, filtering according to requirements.

In the system, the entries go through a life-cycle, see Figure 4. It can be divided in the two interconnected circle for reference modeling on the left side and application modeling on the right side. If there is not yet a reference for an use case, one can be created. First, requirements of the general scenario are analyzed and determined. Besides this, comparable entries with similar requirements have to be identified. These can serve as architecture building blocks for a composition in order to fulfill the deviating conditions. Thereupon either a new version or variant is to be created. The solution approach is to be described with all its peculiarities. The quality of the model is determined by the expertise of the modeler, the chosen notation and the procedure. Especially for compositions, the links with information on interface connections must be traced. Possible improvements are derived from the result of the analysis or feedback, which cause re-designs. As soon as this is completely stored in the system, the approval for further use takes place via a release. This is expressed accordingly by the attribute status. Subsequently, the model can be implemented and used in the domain. Appropriate monitoring is used to control the solution on the basis of metrics. Depending on different domains, parallel variants can be instantiated based on this. Adaptations and improvements are traceable via successor relationships with versioning. Here over the status accordingly the relationship is produced. Over a feedback function comments can be left to specific entry. This serves a sustainable further development and is part of the continuous improvement and maintenance strategy. If the references are no more use due to technological progress, these are to be marked as invalid.

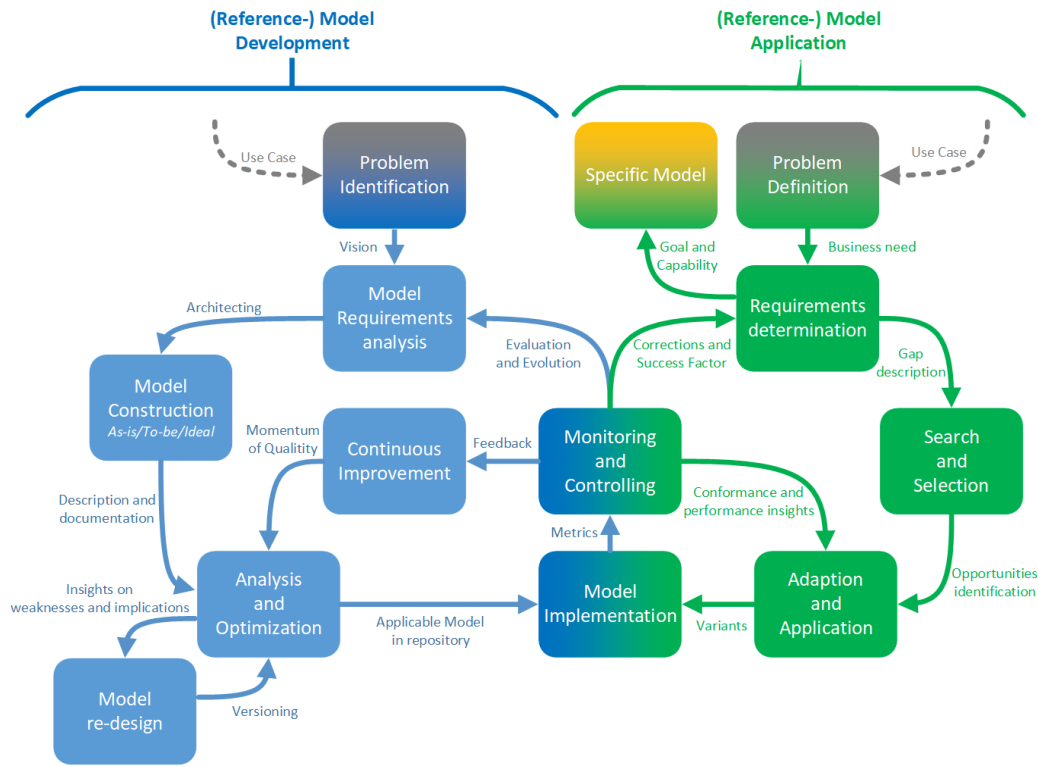

Figure 4. Life-cycle for creating, adapting, and maintaining models in the library.

In order to make the maintenance of models agile, attention must be paid to the documentation of dependencies between individual architectures. If an architecture is revised and modified, this automatically triggers an event to check the linked architectures. This behavior is traced by the attribute status and can create a cascading effect. A supporting system marks the affected areas and informs the responsible authors.

Other methodological suggestions for roles and governance collaboration depend on the specific deployment area. For the frameworks TOGAF and ArchiMate, we refer to the following article [33].

## 5. CASE STUDY AND PRELIMINARY RESULTS

Initial experience was gained in the course of an industrial project. A joint library was created between a smart product provider and an IT service provider with several data centers, in line with the example in Section 2. The companies work in a model-driven manner throughout, so that uniform structures via models can be expected to have a positive impact on operational business.

As test data set, the functional area *Incident Management* is selected according to the frameworks ITIL and IT4IT. It includes: reference models, architecture building blocks, data structures, responsible roles, checklists and many more. These specific models are realized in their own tools depending on their meta-model, which maintains flexibility. The associated models are used as an example to run through the life-cycle in our library and serve as the basis for evaluation.

The first draft for a realization was based on MS-Excel and MS-Visio to obtain a first impression for a deep analysis. It is widely used and the functional scope is powerful, especially in terms of linking data. The tabular listing corresponds to the idea of Fettke and Loos [7], but quickly reaches its limits in terms of clarity and complexity. Figure 5 provides an impression of these prototypes.

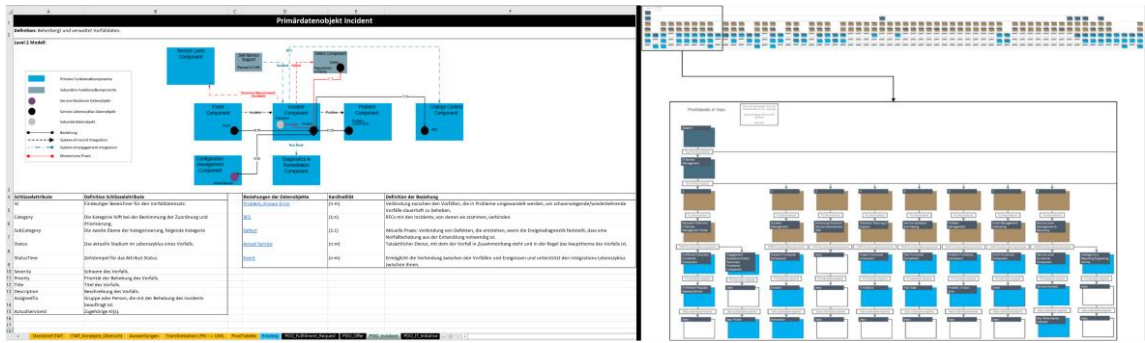

Figure 5. Prototypes in MS-Excel and MS-Visio for first experiences.

In the MS-Excel approach, a worksheet serves as an overview, with each element having an ID for better structuring and linking. The other worksheets are divided according to the attributes *Category* and *Layer*. This structuring allows an extension according to the categories of a domain taxonomy in the width. Here, the flat hierarchy is limited by the software and a missing tree structure. If a model or building block is used in multiple areas, this can be entered accordingly by linking. In this way, all relevant categories can be covered in a consistent and synchronized way. The workflow of usage and life-cycle support is realized manually. The frameworks ITIL and IT4IT categorize the data in over 54 functional units, which we are replicated in our library. Through correlation of models between reference and adapted allows rough statements about the degree of coverage and the fulfillment of specifications. This is done by existence relations with complex expressions in the software suit. Graphical models can be generated automatically in MS-Visio by pivoting the data elements.

Based on the experience of the first two prototypes, an improved software solution for the library as a knowledge base was evaluated. The tools *LexiCan*, *Brain*, and *CherryTree* turned out to be possible alternatives. The implementation in LexiCan can be seen in the Figure 6. It becomes clear that structuring the type layer and domain categories as a tree structure leads to much more clarity.

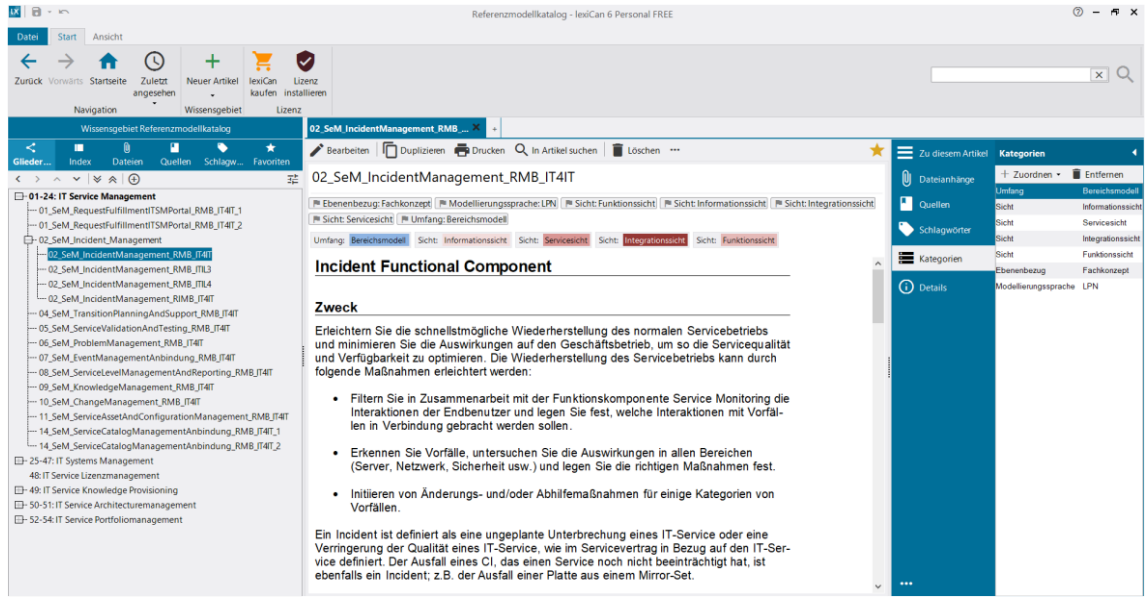

Figure 6. Prototype in the personal Wiki-Software LexiCan.

Nevertheless, also these solutions currently lack possibilities for the direct treatment of models as well as multi-user abilities. In addition, correlations of models and a combination of model components are only indirectly supported. Regardless of this, a clear added value in the daily work with models could already be achieved within the scope of the study.

## 6. CONCLUSIONS

In summary, the present approach describes essential foundations of a model library as a knowledge base with individual adaptability. The focus is on the management of Business-IT-Alignment with reference and application models, but it is not limited to it. Multiple frameworks like TOGAF, NAF, ITIL and IT4IT are adhered. It extends them in detail over multiple level where these frameworks do not provide information. Our case study shows that considerable added value can be achieved using simple standard software or specialized knowledge management tools. Furthermore, it was analyzed to what extent the mentioned standards actually help a modeler in information modeling.

In the future, we extend the library with further reference models and best practices. A next prototype will allow integrated treatment of models in different meta-models via plugins.

In a further expansion stage, the change management of models across several versions and variants is to be made graphically visible. Based on a standardized XML format for models, deviations can be easily identified and highlighted in visual representations. In addition, we investigate relationships and correlations between reference models and individual instances. This allows detailed statements about the degree of coverage and compatibility. Furthermore, an automated determination of reference models from multiple instance models is envisaged.

### ACKNOWLEDGEMENTS

Special thanks to Mr. Lukas Köhler for his support in the development of the basic prototype and his passion for this research.

## AUTHORS

**Peter Hillmann** is a postdoctoral researcher at the Universität der Bundeswehr München, Germany. He received a M.Sc. in Information-System-Technology from Dresden University of Technology (2011) and a Dr. rer. nat. (Ph.D. in science) degree in Computer Science (2018) from the Universität der Bundeswehr München. His areas of research are cyber security, distributed systems as well as enterprise architecture and optimization.

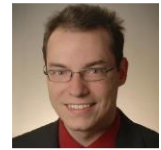

**Diana Schnell** is a research associate in the Department of Applied Computer Science at Universität der Bundeswehr München, Germany. She holds a M.Sc. in Business Administration and Engineering from Universität Duisburg-Essen. Her research interests include enterprise architecture management and an associated improvement in business alignment.

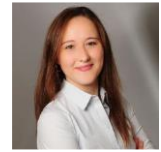

**Harald Hagel** is a senior researcher at the Universität der Bundeswehr München, Germany. He obtained a Dr.-Ing. degree in mechanical engineering (1988) from the Universität der Bundeswehr München. His research area is the application of hybrid and knowledge-based modeling techniques based on business process engineering in industrial practice.

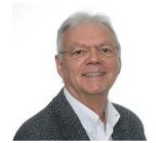

**Andreas Karcher** holds the Chair for Software Tools and Methods for Integrated Applications in the Department of Applied Computer Science at Universität der Bundeswehr München, Germany since 2003. The Chair of Software Tools and Methods for Integrated Applications focuses its research and teaching on digital transformation on the basis of a model- and architecture-based design of integrated application system landscapes.

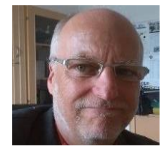